%% file: skeleton.tex
\documentclass{PoS}

\usepackage{amsmath}
\usepackage{amssymb}
\usepackage{subfigure}

\title{Extraction of the isovector magnetic form factor of the nucleon at zero momentum}

\ShortTitle{Extraction of the isovector magnetic form factor of the nucleon at zero momentum}

\author{Constantia Alexandrou\\
        Department of Physics, University of Cyprus, P.O. Box 20537, 1678 Nicosia, Cyprus and\\
        Computation-based Science and Technology Research Center, 20 C. Kavafi Str.,\\
        2121 Nicosia, Cyprus\\
        E-mail: \email{alexand@ucy.ac.cy}}

\author{Martha Constantinou, \\
        Department of Physics, University of Cyprus, P.O. Box 20537, 1678 Nicosia, Cyprus\\
        E-mail: \email{marthac@ucy.ac.cy}}

\author{Giannis Koutsou, \\
        Computation-based Science and Technology Research Center, 20 C. Kavafi Str.,\\
        2121 Nicosia, Cyprus\\
        E-mail: \email{g.koutsou@cyi.ac.cy}}

\author{\speaker{Konstantin Ottnad}\\
        Department of Physics, University of Cyprus, P.O. Box 20537, 1678 Nicosia, Cyprus\\
        E-mail: \email{ottnad@hiskp.uni-bonn.de}}

\author{Marcus Petschlies\\
        Computation-based Science and Technology Research Center, 20 C. Kavafi Str.,\\
        2121 Nicosia, Cyprus\\
        E-mail: \email{m.petschlies@cyi.ac.cy}}

\abstract{The extraction of the magnetic form factor of the nucleon at zero momentum transfer is usually performed by adopting a parametrization for its momentum dependence and fitting the results obtained at finite momenta. We present position space methods that rely on taking the derivative of relevant correlators to extract directly the magnetic form factor at zero momentum without the need to assume a functional form for its momentum dependence. These methods are explored on one ensemble using $N_f=2+1+1$ Wilson twisted mass fermions.}

\FullConference{The 32nd International Symposium on Lattice Field Theory\\
		 23-28 June, 2014\\
		 Columbia University New York, NY}

\input{newcommands}

\begin{document}

\section{Introduction}
We consider the electromagnetic matrix element of the nucleon
\begin{equation}
 \bra{N(p',s')} J_\mu \ket{N(p,s)} = \frac{m_N}{\sqrt{E\l(\vec{p}'\r)E\l(\vec{p}\r)}} \bar{u}(p',s') \l[ \gamma_\mu F_1(q^2) + \frac{i\sigma_{\mu\nu} q_\nu}{2m_N} F_2(q^2) \r] u(p,s) \,,
 \label{eq:matrix_element}
\end{equation}
where $p$, $s$ and $p'$, $s'$ denote momentum and spin of initial and final states, respectively. The momentum transfer squared is given by $q^2=\l(p'-p\r)^2$ and we assume that the final state is produced at rest, hence $\vec{q} = \vec{p'} - \vec{p}  = -\vec{p}$ holds. Assuming $\SU{2}$ isospin symmetry the local electromagnetic current
\begin{equation}
 J_\mu = \frac{2}{3} \bar{u} \g{\mu} u - \frac{1}{3} \bar{d} \g{\mu} d \,,
 \label{eq:elmag_current}
\end{equation}
satisfies the relation
\begin{equation}
 \bra{p} J_\mu \ket{p} - \bra{n} J_\mu \ket{n} = \bra{p} \bar{u} \g{\mu} u - \bar{d} \g{\mu} d \ket{p} \equiv \bra{p} J^\mathrm{iso}_\mu \ket{p} \,,
 \label{eq:isospin_current}
\end{equation}
that defines  the isovector electromagnetic current $J^\mathrm{iso}_\mu$. The use of $J^\mathrm{iso}_\mu$ has the advantage that no disconnected diagrams contribute in the lattice QCD evaluation of the matrix element given in Eq.~(\ref{eq:matrix_element}). Furthermore, we replace the local current by its lattice conserved version avoiding any renormalization factors in our calculations. \par

In Euclidean space-time the momentum transfer squared is given by $Q^2=-q^2$ and the corresponding Dirac and Pauli form factors $F_1\l(Q^2\r)$ and $F_2\l(Q^2\r)$ are related to the (isovector) electric and magnetic Sachs form factors by
\begin{align}
 G_E\l(Q^2\r) &= F_1\l(Q^2\r) - \frac{Q^2}{4m_N^2} F_2\l(Q^2\r) \,, \notag \\
 G_M\l(Q^2\r) &= F_1\l(Q^2\r) + F_2\l(Q^2\r) \,.
\end{align}
On the lattice we compute the gauge-average $\l<.\r>$ of spin-projected two- and three-point functions $C_{2pt}(t,\vec{q})=C_{2pt}(t,\vec{q},\Gamma_0)$, $C^\mu_{3pt}(t,\vec{q},\Gamma^\nu)$ and build the optimized ratio
\begin{equation}
 R_\mu(t_s,t,\vec{q},\Gamma_\nu) = \frac{\bigl<C_\mu^{3pt}(t_s,t,\vec{q},\Gamma_\nu)\bigr>}{\bigl<C^{2pt}(t_s,\vec{0})\bigr>} \sqrt{\frac{\bigl<C^{2pt}(t_s-t,\vec{q})\bigr> \bigl<C^{2pt}(t, \vec{0})\bigr> \bigl<C^{2pt}(t_s, \vec{0})\bigr>}{\bigl<C^{2pt}(t_s-t, \vec{0})\bigr> \bigl<C^{2pt}(t, \vec{q})\bigr> \bigl<C^{2pt}(\vec{q}, t_s)\bigr>}} \,,
 \label{eq:ratio}
\end{equation}
where $t_s$, $t$ denote the (fixed) sink and (running) insertion timeslices, respectively. The source timeslice is chosen to be zero. The three-point functions are computed using sequential inversions through the sink \cite{Dolgov:2002zm} in order to obtain the full $Q^2$--dependence.  For large Euclidean times $t$ and $t_s-t$, the ground state dominates the ratio and $R^\mu(t_s,t,\vec{q},\Gamma^\nu)$ approaches a plateau, i.e.
\begin{equation}
 \lim\limits_{t\rightarrow\infty} \ \lim\limits_{t_s-t\rightarrow\infty} R^\mu(t_s,t,\vec{q},\Gamma^\nu) = \Pi^\mu\l(\vec{q}, \Gamma^\nu\r) \,,
\end{equation}
which allows us to extract the isovector Sachs form factors employing an appropriate choice of projectors and insertion indices
\begin{align}
 \Pi_0\l(\vec{q},\Gamma_0\r) &= -C\frac{E\l(\vec{q}\r)+m_N}{2m_N} G_E\l(Q^2\r) \label{eq:G_E} \,, \\
 \Pi_i\l(\vec{q},\Gamma_0\r) &= -C\frac{i}{2m_N} q_i G_E\l(Q^2\r) \label{eq:q_G_E} \,, \\
 \Pi_i\l(\vec{q},\Gamma^k\r) &= -C\frac{1}{4m_N} \epsilon_{ijk} q_j G_M\l(Q^2\r) \label{eq:q_G_M} \,,
\end{align}
where the relevant projectors are given by $\Gamma_0 = \frac{1}{2} \l(1+\g{0}\r)$, $\Gamma_k = \frac{1}{4} \Gamma_0 i \g{5}\g{k}$ and $C=\sqrt{\frac{2m_N^2}{E\l(\vec{q}\r)\l(E\l(\vec{q}\r)+m_N\r)}}$. From Eq.~(\ref{eq:G_E}) it is obvious that the isovector electric moment $G_E\l(0\r)=1$ can be extracted directly in a lattice calculation, whereas for the case of the anomalous magnetic moment $G_M\l(0\r)$ no relation without a multiplicative momentum factor exists. The standard method to obtain an estimate for $G_M\l(0\r)$ is to choose a fit ansatz that describes the data at non-zero momentum transfer and use the fitted parameters to extrapolate to zero-momentum. However, this introduces a model dependence, which given the discrete nature of $Q^2$ in a lattice calculation can be problematic. Here we  follow a different, model-independent approach that employs correlation functions in position space.\par

\section{Position space methods (I)}
Assuming continuous momenta one can formally isolate $G_M\l(0\r)$ from Eq.~(\ref{eq:q_G_M}) by applying a derivative with respect to $q_j$
\begin{equation}
 \lim\limits_{q^2\rightarrow0} \frac{\partial}{\partial q_j} \Pi_i\l(\vec{q}, \Gamma_k\r) =  \frac{1}{4m_N} \, \epsilon_{ijk} G_M\l(0\r) \,.
 \label{eq:derivative}
\end{equation}
On the lattice a corresponding procedure to remove the factor $q_j$ can be defined in different ways. One possibility is given by the formal application of a continuum-like derivative to the ratio in Eq.~(\ref{eq:ratio})
\begin{align}
 \lim\limits_{q^2\rightarrow0} \frac{\partial}{\partial q_j} R_i(t_s,t,\vec{q},\Gamma_k) &= \lim\limits_{q^2\rightarrow0} \frac{\bigl<\frac{\partial}{\partial
  q_j}C_i^{3pt}(t,\vec{q},\Gamma_k)\bigr>}{\bigl<C^{2pt}(t_s,\vec{0})\bigr>} \notag \\
  &=  \lim\limits_{L\rightarrow\infty}\frac{1}{\bigl<C^{2pt}(t_s,\vec{0})\bigr>} \cdot \bigl<\sum\limits_{x=-L/2+a}^{L/2-a} i x_j C_i^{3pt}(t, \vec{x})\bigr> \,,
 \label{eq:continuum_derivative}
\end{align}
where in the second line the three-point function $C_i^{3pt}(t, \vec{x})$ in position space has been introduced. Note that any derivatives of two-point functions in the above expression vanish exactly. In finite volume this expression approximates the derivative of a $\delta$-distribution in momentum space,
\begin{equation}
a^3\sum\limits_{\vec{x}} i x_j C_i^{3pt}(t; \vec{x}) = \frac{1}{V}\sum\limits_{\vec{k}} \l(a^3\sum\limits_{\vec{x}} i x_j \exp(i\vec{k}\cdot\vec{x})\r) C_i^{3pt}(t, \vec{k})  \xrightarrow{L\rightarrow\infty}\frac{1}{(2\pi)^3}\int d^3\vec{k} \frac{\partial}{\partial k_j}\delta^{(3)}(\vec{k})  C_i^{3pt}(t, \vec{k}) \,,
 \label{eq:delta_approximation}
\end{equation}
which implies a residual $t$--dependence $C_i^{3pt}(t,\vec{q},\Gamma_k) \sim \exp(-\Delta E t)$, where $\Delta E = E(\vec{q}) -m_N$ is the momentum transfer between final and initial state and $\Delta E\rightarrow 0$ for $L\rightarrow \infty$. \par

The basic building blocks for this method are the standard two-point functions and the continuum derivative-like three-point functions $\l<\frac{\partial}{\partial q_j}C_i^{3pt}(t,\vec{q},\Gamma_k)\r>$. In an actual lattice calculation the latter involve the computation of the full three-point function in position space before performing the multiplication by $x_j$ in the final Fourier transform and building the ratio in Eq.~(\ref{eq:continuum_derivative}). Moreover, this method requires a sufficiently large cutoff for the summation in Eq.~(\ref{eq:delta_approximation}), which needs to be checked explicitly on a given gauge ensemble. We will refer to this approach as the \emph{continuum derivative} method. \par

\section{Position space methods (II)}
A second method is obtained by starting with a fit to the plateau in Eq.~(\ref{eq:ratio}), hence removing any time dependence. Let us first consider the case of on-axis momenta, e.g. $\vec{q}=(\pm q, 0,0)^T$ (and all permutations thereof). Before we apply the fit to the plateau, we average over all momentum directions and contributing index combinations according to Eq.~(\ref{eq:q_G_M}) for a given $q$-value, and we denote the corresponding fitted ratios by $\Pi(q)$. \par
In a next step we perform a Fourier transform to obtain a ratio $\Pi(y)$ in position space for which $\Pi(y) \approx -\Pi(-y)$ holds up to statistical fluctuations. Note that in practice this requires a cutoff $q_\mathrm{max}$ in the Fourier transform as the original ratio can only be calculated for a limited number of lattice momenta in an actual simulation. With $n=y/a$ we have
\begin{equation}
 \Pi(y)=\l\{\begin{array}{ll}
   +\Pi(n), & n=0,...,N/2 \\
   -\Pi(N-n), & n=N/2+1,...,N-1, \ N=L/a
   \end{array}\r. \,,
\end{equation}
and averaging over positive and negative values of $y$ we obtain an exactly antisymmetric expression $\overline{\Pi}(n)$. Finally, $\overline{\Pi}(n)$ is transformed back in a way that allows us to introduce continuous momenta. Starting from
\begin{align} 
 \Pi(k) &= \l[\exp(ikn)\overline{\Pi}(n)\r]_{n=0,\,N/2} + \sum\limits_{n=1}^{N/2-1}\exp(ikn)\overline{\Pi}(n) + \sum\limits_{n=N-1}^{N/2+1}\exp(ik(N-n))\overline{\Pi}(n) \notag \\
        &=\l[\exp(ikn)\overline{\Pi}(n)\r]_{n=0,\,N/2} + 2i \sum\limits_{n=1}^{N/2-1} \overline{\Pi}(n) \sin\l(\frac{k}{2}\cdot (2n) \r) \,
\end{align}
and defining $\hat{k} \equiv 2\sin\bigl(\frac{k}{2}\bigr)$ and $P_n\bigl(\hat{k}^2\bigr)=P_n\bigl(\bigl(2\sin\bigl(\frac{k}{2}\bigr)\bigr)^2\bigr) = \sin(nk) / \sin\bigl(\frac{k}{2}\bigr)$ we have
\begin{equation}
 \Pi(\hat{k})-\Pi(0) = i \sum\limits_{n=1}^{N/2-1} \hat{k}\, P_n \,\bigl(\hat{k}^2\bigr) \overline{\Pi}(n) \,.
\end{equation}
Note that $P_n \,\bigl(\hat{k}^2\bigr)$ can be related to Chebyshev polynomials of the second kind and is hence analytic in $(-\infty, +1)$, allowing for an evaluation of $\Pi(\hat{k})$ at any intermediate value. In the above expression we can divide by $\hat{k}$, and -- including the appropriate kinematic factors for $\Pi(n)$ -- we obtain the desired expression for the nucleon magnetic moment without explicit momentum factors
\begin{equation}
 G_M(\hat{k}^2) = i\sum\limits_{n=1}^{N/2-1} P_n(\hat{k}^2) \, \overline{\Pi}(n) \,.
 \label{eq:method2}
\end{equation}
Note that in the limit $\hat{k}\rightarrow0$ division by $\hat{k}$ is equivalent to applying a derivative with respect to $\hat{k}$. We remark that it is straightforward to extend this method to arbitrary off-axis momentum classes $M(q_\mathrm{off}^2) = \l\{\vec{q} \ | \ \vec{q}=\{\pm q, q_1, q_2\} \,, \ q_1^2+q_2^2=q_\mathrm{off}^2\r\}$, where $\{\pm q, q_1, q_2\}$ denotes all permutations of $\pm q$, $q_1$ and $q_2$, with $ q=2\pi n/L$ for $n=0,...,N/2$. However, to combine the results for $G_M(Q^2)$ for different $q_\mathrm{off}^2$--classes as a function of (now continuous) Euclidean momenta $Q^2=Q^2(\hat{k}, q_\mathrm{off}^2)$ we need to consider an analytic continuation for classes with $q_\mathrm{off}^2>0$ to reach zero momentum, i.e. $Q^2=0$. This is achieved by consistently replacing $k\rightarrow i\kappa$ and $\hat{k} \rightarrow i\hat{\kappa} = -2\sinh\bigl(\frac{\kappa}{2}\bigr)$ in the derivation outlined above. Note that in this case one has $P_n\bigl(\hat{\kappa}^2\bigr) = \sinh(n\kappa) / \sinh\bigl(\frac{\kappa}{2}\bigr)$. Similar approaches using analytic continuation have been used in the context of calculating hadronic vacuum polarizations \cite{Bernecker:2011gh,Feng:2013xsa,Feng:2013xqa}. In the following we will refer to this approach as the \emph{y--summation} method. \par

\section{Lattice results}
We have tested the aforementioned methods on one ensemble of $N_f=2+1+1$ Wilson twisted mass fermions corresponding to a charged pion mass of $M_\mathrm{PS} = 373 \> \mathrm{GeV}$ generated by the ETM collaboration \cite{Baron:2010bv,Baron:2010th,Baron:2011sf}. In the notation of \cite{Baron:2011sf} this ensemble is denoted by $B55.32$ and we refer to \cite{Baron:2011sf} for further details on the simulation. For the calculation of the three-point functions using the sequential method we restrict ourselves to source-sink separations of $t_\mathrm{sep}=12$, which turns out large enough to avoid excited state contaminations for the observables in question. All errors have been calculated from a jackknife analysis. \par

\begin{figure}[t]
  \centering
  \subfigure[]{\includegraphics[width=.48\linewidth]
    {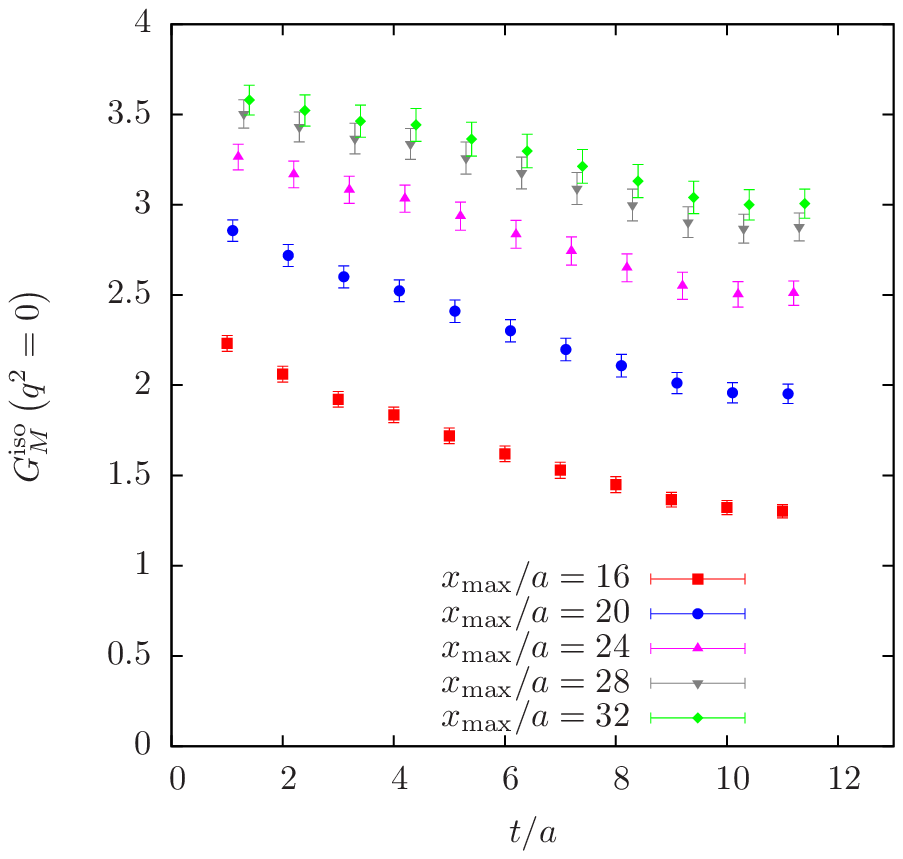}}\quad
  \subfigure[]{\includegraphics[width=.48\linewidth]
    {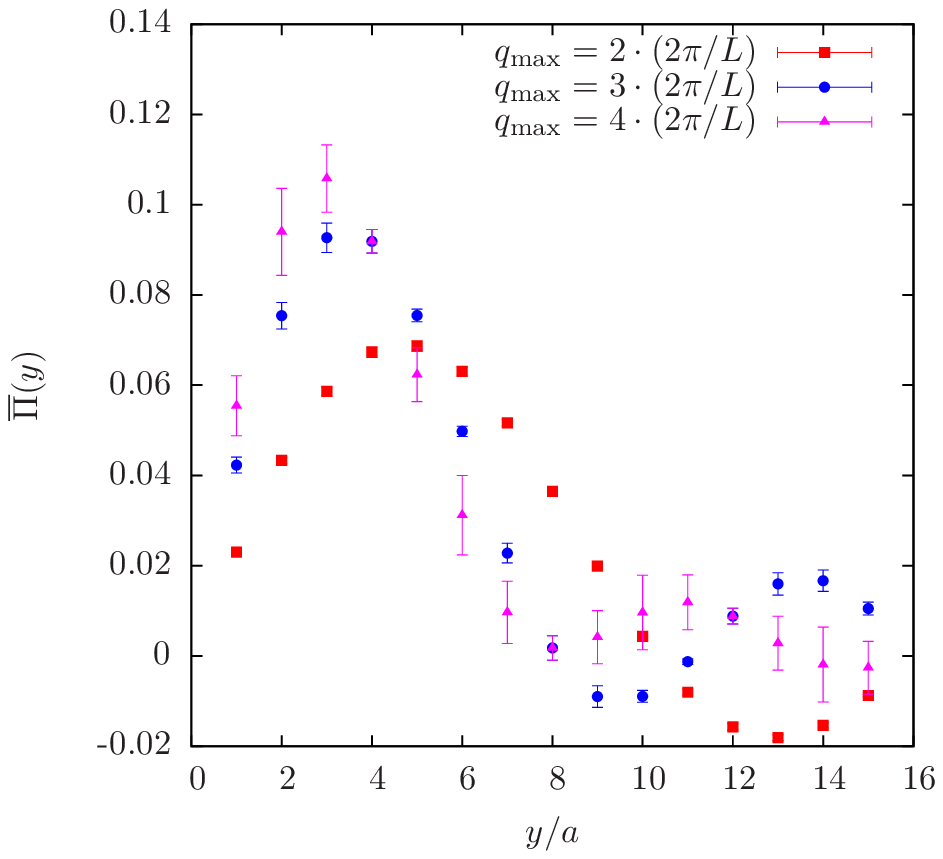}}\quad
  \caption{(a) The (t-dependent) isovector magnetic moment as derived in Eqs.~(\protect\ref{eq:derivative})--(\protect\ref{eq:delta_approximation}) using several cutoff values for the $x$-summation. (b) The position space ratio $\overline{\Pi}(y)$ as a function of the momentum summation cutoff $q_\mathrm{max}$.}
  \label{fig:tests}
\end{figure}

In the left panel of Fig.~\ref{fig:tests} we show the residual $t$-dependence of the ratio for $G_M^\mathrm{iso}(0)$ for different cutoff values $x_\mathrm{max}$ of the $x$--summation in Eq.~(\ref{eq:continuum_derivative}) using 300 gauge configurations. It is obvious that no actual plateau is reached within errors for any of the available cutoff values, although larger values tend to give a more flat behavior, as expected. Clearly, a larger value of $L/a$ would be required to allow for a meaningful extraction of the magnetic moment. The resulting values for the magnetic moment lie all well below the one extracted from a standard dipole fit ansatz using 1200 configurations of the same gauge ensemble in \cite{Alexandrou:2013joa}, which gave $G_M^\mathrm{iso}=3.93(12)$. \par

For the y--summation method we employed 1560 gauge configurations of the same ensemble, each separated by four Monte-Carlo trajectories. The errors are calculated from a jackknife analysis with binning, although it turns out that autocorrelation is negligible for the isovector magnetic moment. In the right panel of Fig.~\ref{fig:tests} we show the dependence of the position space ratio $\overline{\Pi}(y)$ for the on-axis case on the choice of the momentum cutoff $q_\mathrm{max}$ for the Fourier transform. From the behavior at larger values of $y/a$, we conclude that contributions from momenta $q>4\cdot(2\pi / L)$ should be negligible. \par

\begin{figure}[t]
 \centering
 \subfigure[]{\includegraphics[width=.48\linewidth]
  {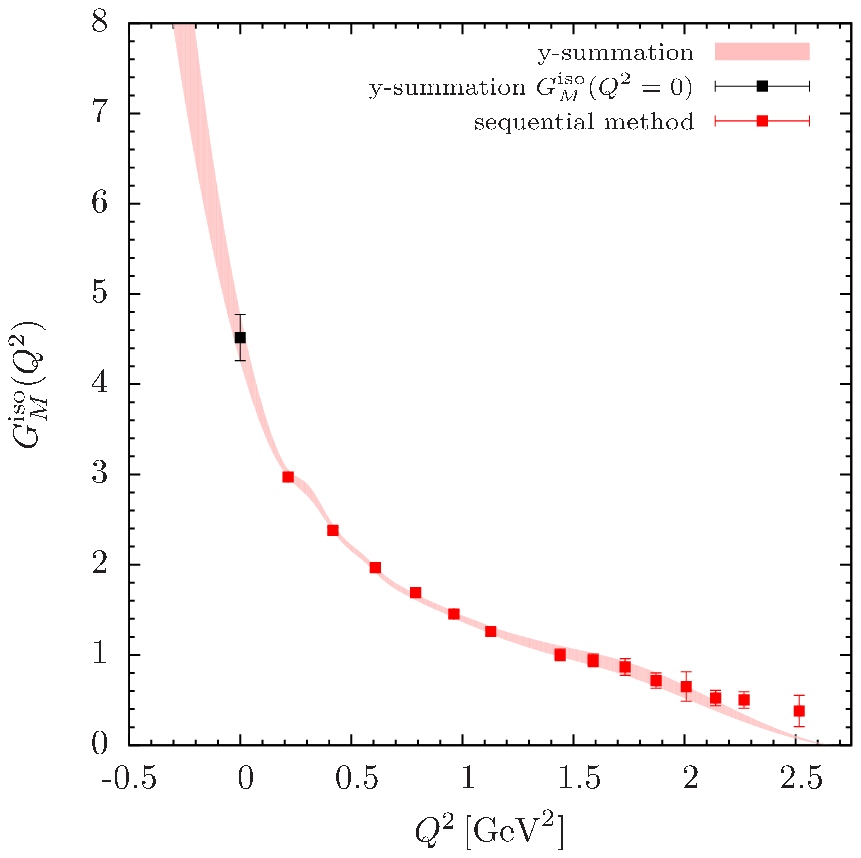}}\quad
 \subfigure[]{\includegraphics[width=.48\linewidth]
  {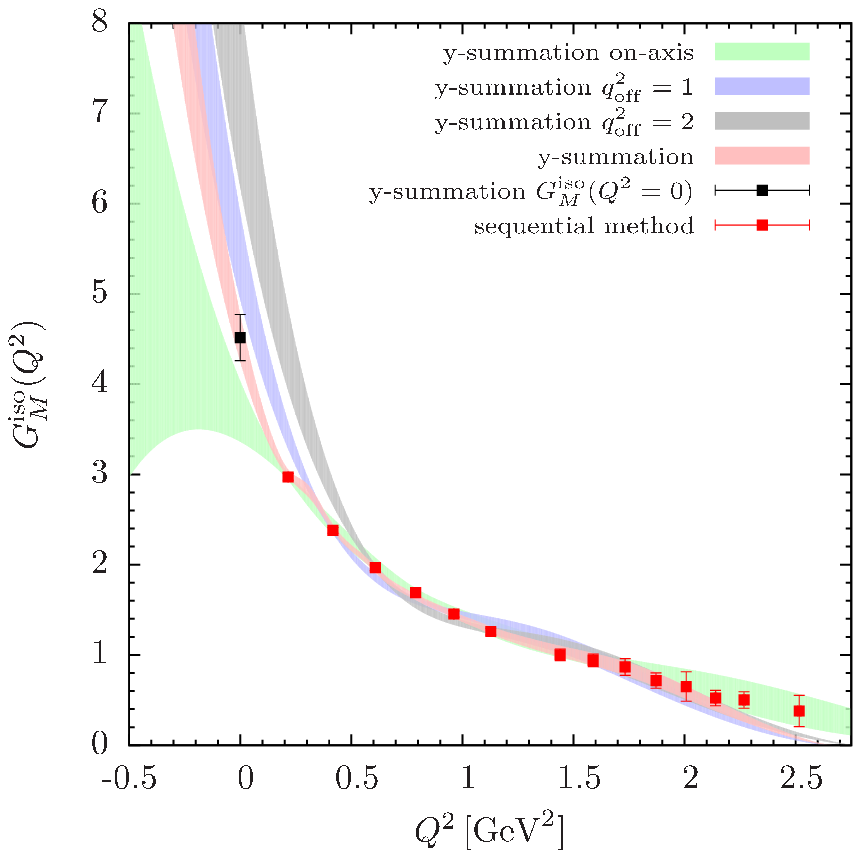}}\quad
 \caption{(a) Isovector $G_M(Q^2)$ extracted using the y--summation for the $B55.32$ ensemble. The red band is the weighted averaged over the five lowest momentum classes used for y--summation and the resulting value of the magnetic moment is shown by a black square. The red squares correspond to the standard sequential method. (b) The same, but additionally showing bands for the first three momentum classes separately.}
 \label{fig:results}
\end{figure}

The left panel of Fig.~\ref{fig:results} shows our main result for the isovector magnetic form factor. For this plot we have applied the y--summation method for the first five momentum classes, i.e. $M(q_\mathrm{off}^2\leq5)$\footnote{Note that here as well as in the following discussion we have suppressed all factors of $(2\pi/L)^2$ in the values of $q_\mathrm{off}^2$ for better readability.} and we have used a value of $a=0.082fm$ for the lattice spacing \cite{Alexandrou:2013joa} to convert to physical units. The red $1\sigma$-band is obtained as the error-weighted average of the results over the five classes. However, at least for low values of $Q^2$ the first two classes clearly dominate the resulting signal as off-axis classes with $q_\mathrm{off}^2\geq2$ have much larger errors. \par

For the isovector magnetic moment we obtain $G_M^\mathrm{iso}=4.51(26)$, which is compatible with the experimental value of $4.7$~\cite{Agashe:2014kda} but larger than the aforementioned result from a dipole fit. In order to investigate this difference further we have shown the resulting bands from the first three classes $M(q_\mathrm{off}^2\leq2)$ in the right panel of Fig.~\ref{fig:results} separately. To guide the eye we have also again included the weighted average over the lowest five off-axis classes by a red band, denoted by "y--summation" only. Looking at the two bands for the on-axis class and the first off-axis class it becomes obvious that there are rather large lattice artifacts present at least at small values of $Q^2$. This is also the reason for the kink in the red band between the two red points corresponding to the lowest $Q^2$--values, because the splitting between $M(q_\mathrm{off}^2=0)$ and $M(q_\mathrm{off}^2=1)$ is sizable while the respective errors behave differently for the two classes. The latter is caused by the fact that the data point at lowest $Q^2$ contributes only to the calculation of the band for $M(q_\mathrm{off}^2=0)$ and the second point in $Q^2$ only to $M(q_\mathrm{off}^2=1)$. Note that the next off-axis class $M(q_\mathrm{off}^2=2)$ appears to give even larger values at small $Q^2$, but its errors are already so large that there is almost no contribution to the red band coming from this class at such small momenta. We have also checked that even higher classes give compatible values (within their very large errors). \par

\section{Summary}
In this work we have shown that a position space method (i.e. y--summation) together with analytic continuation for off-axis momentum classes allows for a model-independent extraction of the isovector magnetic moment. The slightly larger value compared to a previous calculation from a fit can be traced back to different lattice artifacts coming from each contribution of the included off-axis momentum classes. In the standard approach this is masked by simply averaging over all values belonging to a common $Q^2$--value before applying the dipole fit ansatz to the resulting data for $G_M^\mathrm{iso}(Q^2)$, leading to different artifacts for the fitted values. We remark that the value $G_M^\mathrm{iso}=4.51(26)$ obtained from the y--summation method is somewhat closer to the experimental value, but from the currently available data we cannot decide whether lattice artifacts are larger for the y--summation method or the standard fit ansatz. It would be useful to have the continuum derivative method as a third approach, however, this would require a larger spatial extend of the lattice and / or a fit to deal with the residual t-dependence in order to extract a meaningful result. \par

From a technical point of view, the y--summation method is applicable to any form factor in its domain of analyticity, regardless of its kinematic prefactor in the decomposition. In particular it can be used to extract the pseudo-scalar form factor $G_p(0)$ involved in the nucleon matrix element of the axial-vector charge. This is not so well measured as $G_M(0)$ and can provide useful input to phenomenology.

\acknowledgments{We would like to thank all members of ETMC for the most enjoyable collaboration. Numerical calculations have used HPC resources from John von Neumann-Institute for Computing on the JUQUEEN and JUROPA systems at the research center in J\"ulich. Additional computational resources were provided by the Cy-Tera machine at The Cyprus Institute funded by the Cyprus Research Promotion Foundation (RPF), NEAY$\Pi$O$\Delta$OMH/$\Sigma$TPATH/0308/31. M.C. is supported by the Cyprus Research Promotion Foundation under the contract TECHNOLOGY/$\Theta$E$\Pi$I$\Sigma$/0311(BE)/16.}

\bibliographystyle{h-physrev5}
\bibliography{bibliography}

\end{document}

%% file: newcommands.tex
\newcommand{\g}[1]{\gamma_{#1}} 
\renewcommand{\l}{\left}
\renewcommand{\r}{\right}

\newcommand{\bra}[1]{\left< #1 \right|} 
\newcommand{\ket}[1]{\left| #1 \right>} 

\newcommand{\SU}[1]{\mathrm{SU}\l(#1\r)}
